# Synthesis of narrow SnTe nanowires using alloy nanoparticles

*Pengzi Liu[1,2], Hyeuk Jin Han[1,2], Julia Wei[3], David J. Hynek[1,2], James L. Hart[1,2], Myung Geun Han[4], Christie J. Trimble[5,6], James R. Williams[5,6,7], Yimei Zhu[4], Judy J. Cha[1,2,8]\**

[1] Department of Mechanical Engineering and Materials Science, Yale University, New Haven, CT 06511, USA

[2] Energy Sciences Institute, Yale West Campus, West Haven, CT 06516, USA

[3] Department of Physics, Yale University, New Haven, CT 06511, USA

[4] Department of Condensed Matter Physics and Materials Science, Brookhaven National Laboratory, Upton, New York 11973, United States.

[5] Department of Physics, University of Maryland, College Park, MD 20742, USA

[6] Joint Quantum Institute, University of Maryland, College Park, MD 20742, USA

[7] Quantum Materials Center, University of Maryland, College Park, MD 20742, USA

[8] Canadian Institute for Advanced Research Azrieli Global Scholar, Canada

**ABSTRACT**

Topological crystalline insulator tin telluride (SnTe) provides a rich playground to examine interactions of correlated electronic states, such as ferroelectricity, topological surface states, and superconductivity. Making SnTe into nanowires further induces novel electronic states due to one-dimensional (1D) confinement effects. Thus, for transport measurements, SnTe nanowires must be made narrow in their diameters to ensure the 1D confinement and phase coherence of the topological surface electrons. This study reports a facile growth method to produce narrow SnTe nanowires with a high yield using alloy nanoparticles as growth catalysts. The average diameter of the SnTe nanowires grown using the alloy nanoparticles is 85 nm, nearly a factor of three reduction from the previous average diameter of 240 nm using gold nanoparticles as growth catalysts. Transport measurements reveal the effect of the nanowire diameter on the residual resistance ratio and magnetoresistance. Particularly, the ferroelectric transition temperature for SnTe is observed to change systematically with the nanowire diameter. *In situ* cryogenic cooling of narrow SnTe nanowires in a transmission electron microscope directly reveals the cubic to rhombohedral structural transition, which is associated with the ferroelectric transition. Thus, these narrow SnTe nanowires represent a model system to study electronic states arising from the 1D confinement, such as 1D topological superconductivity as well as a potential multi-band superconductivity.

**INTRODUCTION**

Making topological materials into nanostructures is advantageous to enhance the contribution of the topological surface states in transport measurements, as evident in Aharonov-Bohm oscillation studies of the surface states in topological insulator nanowires[1-4] and nanoribbons[5-6] and exotic quantum Hall oscillations observed in nanoslabs of Dirac semimetals.[7] Moreover, novel electronic states can emerge at the nanoscale, such as the Majorana bound states expected in 1D topological superconductors.[8-12] In topological crystalline insulator SnTe,[13-14] the nanoscale confinement can have a profound impact on the nature of ferroelectric domains,[15] electron coupling between topological surface states,[16] and superconductivity.[17-20] Several transport studies of SnTe nanostructures highlight the effects of the nanoscale confinement.[19, 21-22] For example, the proximity-induced superconductivity in SnTe nanowires is recently suggested to have a novel pairing symmetry of $s \pm is$, which was observed in nanowires with diameters < 250 nm, and the phenomenon was more pronounced for smaller diameter nanowires.[18] Thus, it is critical to develop precision synthesis for topological nanomaterials with controlled dimensions, morphology, and crystalline quality.

Most of the topological nanomaterials studied to date have been synthesized using metal nanoparticle-catalyzed chemical vapor deposition (CVD),[5, 11, 22-29] in which the vapor-liquid-solid (VLS) growth and vapor-solid (VS) growth often occur concurrently to produce nanoribbons and nanoplates with a broad distribution of their widths, lacking control on the dimensions and morphology of the nanostructures. If the growth is strictly VLS, nanowires should be produced with their diameters identical to the sizes of the metal nanoparticles used for growth.[30-31] It is possible to tune the growth conditions carefully to promote VLS growth over the VS growth in order to control the morphology of the nanowires, which has been exploited to grow heterostructure semiconducting nanowires with the heterostructure segments along the axial and radial directions.[32-33] For topological nanomaterials however, such careful growth studies are sparse. For synthesis of SnTe nanostructures, several works have explored the effects of substrates, catalyst particles, and chemical potentials of gaseous species on the morphologies of SnTe nanostructures,[26, 34-37] but most SnTe nanowires reported to date have large diameters. Another route to promote VLS growth for narrow nanowires is to tune the composition of the metal nanoparticles to satisfy the supersaturation conditions more easily and lower the nucleation barrier for the solid phase.[38-42] In this work, we use Au-Sn-Te alloy nanoparticles as growth catalysts to grow narrow SnTe nanowires with a high yield. The average diameter of the SnTe nanowires grown using the alloy nanoparticles is 85 nm, in contrast to 240 nm using the gold nanoparticles. We also observe higher ferroelectric transition temperatures for narrower nanowires. *In situ* cryogenic transmission electron microscopy (cryo-TEM) experiments confirm the expected structural transition from the room temperature cubic phase to the low temperature rhombohedral phase for these SnTe nanowires. Thus, the narrow SnTe nanowires provide a rich playground to examine the possible multi-band superconductivity, Majorana bound states, and interplay between ferroelectricity and topological orders.

**RESULTS AND DISCUSSION**

The alloy nanoparticles were first produced by reacting Au nanoparticles with SnTe vapor at elevated temperatures in a horizontal tube furnace with a single temperature zone (Lindberg/Blue M, Thermo Fisher Scientific). A solution containing 20 nm Au nanoparticles (Sigma Aldrich, Prod. #741965) was drop-casted on the $SiO_x$/Si substrates that were pre-treated with a poly-L-Lysine solution (Sigma Aldrich, 0.01%). The substrates were blow-dried with nitrogen gas for 2 minutes and placed at downstream inside a quartz tube with 1-inch diameter at distances 9.5 cm to 12.5 cm away from the center of the furnace. 0.1 g SnTe source powder (Sigma Aldrich, 99.999%) was finely ground and placed in a quartz boat, which was then loaded at the center of the quartz tube. The quartz tube was purged with argon gas (99.999%) 4 times and held at 600℃ for 1 hour after ramping from room temperature within 30 minutes. The pressure was maintained at 2 Torr with a constant Ar flow at the rate of 20 sccm. After the reaction, the furnace was cooled down to room temperature with the fast-cooling process as described in our previous work.[27]

    Figure 1a shows a scanning electron microscope (SEM) image of the alloy nanoparticles after the anneal, which shows many particles with diameter < 100 nm. A TEM image of one of the particles shows a set of lattice fringes (Figure 1b), and a fast Fourier transform of the TEM

image shows diffraction spots that belong to either Au or AuSn (Figure 1c). Energy dispersive X-ray (EDX) was performed to analyze composition of the alloy nanoparticles, which shows presence of Au, Sn, and Te (Figure 1d). The annular dark field scanning TEM (ADF-STEM) image of the alloy nanoparticles (Figure 1d inset) shows bright and dark regions, which suggests phase separation. The dark region contains slightly more Sn and Te than the bright region although both regions contain Au, Sn, and Te. According to the Au-Sn-Te phase diagram (Supplementary Figure S1), Au, AuSn, SnTe, and $AuTe_2$ are possible at room temperature. While the presence of Te could be due to nucleation of SnTe or $AuTe_2$, formation of SnTe is more likely than formation of $AuTe_2$ based on the phase diagram. Thus, we conclude that the nanoparticles are no longer pure Au nanoparticles, but contain Au and AuSn phases, as well as SnTe nuclei. This conclusion agrees with the observation that the Sn/Te ratios from the nanoparticles are higher than that of SnTe (Figure 1e), indicating presence of SnTe and AuSn phases. The alloy composition of the nanoparticles changes with the growth substrate temperature (Supplementary Figure S2), which suggests varying fractions of Au, AuSn, and SnTe regions in the alloy nanoparticles as a function of substrate temperature. Our findings are in agreement with a previous report, which shows that the Au nanoparticles become alloyed with Sn during synthesis of SnTe nanowires.[35] We note that this reaction is essentially the metal-catalyzed CVD growth of SnTe nanowires,[25-27, 34, 37] and indeed we observe SnTe nanowires and nanoribbons of varying diameters and SnTe microcrystals in addition to the alloy nanoparticles.

For the synthesis of narrow SnTe nanowires, the $SiO_x$/Si substrates decorated with the alloy nanoparticles were loaded into a clean quartz tube at distances from 12.5 cm to 15.5 cm away from the center, along with 0.1 g SnTe source powder placed at the center of the furnace. The temperature profile, growth time, pressure, and the Ar flow rate were kept the same as the first reaction. Narrow SnTe nanowires were found from the substrates placed at 14.5 and 15.5 cm away from the center, which were approximately at 436 and 370℃ during growth, respectively (see Supplementary Figure S3 for the temperature profile of the furnace). Figure 2 shows the comparison of the dimensions and the growth yield of the SnTe nanowires grown using the Au nanoparticles versus the alloy nanoparticles. Figure 2a and 2b show SnTe nanowires grown using the 20 nm Au nanoparticles, and Figure 2c and 2d show SnTe nanowires grown using the alloy nanoparticles. The nanowire yield is significantly higher using the alloy nanoparticles. In addition, the nanowires grown with the alloy particles are much narrower and shorter than the wires grown with the Au nanoparticles. The alloy nanoparticles were observed at the end of the SnTe nanowires, which indicates that they catalyzed the nucleation of the nanowires (Figure 2d). Additionally, the diameter of the nanowire is similar to the diameter of the alloy nanoparticle. By contrast, the SnTe nanowire grown using the Au nanoparticle is much larger than the Au particle (Figure 2b). Figure 2e shows the summary of the comparison between the two cases. Using the alloy nanoparticles, the average diameter of the SnTe nanowires is 85 nm with a standard deviation of 26 nm, while it is 240 nm with a standard deviation of 102 nm using the 20 nm Au nanoparticles. The distribution of the nanowire diameter is significantly narrower and the diameter is reduced by more than a factor of two.

The crystal structure of the narrow SnTe nanowires grown using the alloy nanoparticles was characterized by TEM. Figure 2f-g show TEM images of a 50 nm-wide SnTe nanowire, which shows the expected cubic symmetry and lattice spacing of 3.15Å at room temperature. The Sn/Te ratio of the narrow SnTe nanowire is identical to that of SnTe nanowires grown using Au nanoparticles based on the TEM-EDX data (Figure 2h). While structural defects, such as surface stacking, strain, and vacancies, would exist in these nanowires, the crystalline quality of the nanowires is sufficient to show transport properties that stem from the topological surface states, such as the recent report of novel superconductivity observed in Josephson junctions using these narrow SnTe nanowires as weak links.[18]

The observed synthesis results can be attributed to the lowered nucleation barrier for the precipitation of the SnTe solid phase. Because the alloy particles already contain Sn and SnTe nuclei, the supersaturation conditions to nucleate a solid phase out of the metal catalyst can be reached more easily with the alloy nanoparticle than with the Au nanoparticle.[32] The reduced nucleation barrier thus would promote VLS growth over the side growth via the direct VS deposition, which can explain the small diameter of the SnTe nanowire. Intriguingly, the length of the nanowires got shortened significantly using the alloy nanoparticles, suggesting suppressed growth rates. The growth rate of a nanowire during a VLS growth is a complex function of growth parameters, such as the diffusivities and diffusion lengths of the precursor species, the rate of direct impingements of the precursor species at the metal catalysts, nanowire diameter, and spacings between the nanowires.[43-48] Detailed growth studies of GaAs,[49-50] InAs,[51-52] and GaP[53-54] nanowires highlight the complexity of the growth dynamics. We conjecture that due to the high density of the SnTe nanowires grown on the substrate, growth by direct impingements at the alloy metal particles is unlikely and the nanowire growth stops when the diffusion lengths of the precursors from the substrates are reached. Systematic growth studies must be carried out to fully understand the suppressed growth rate of the SnTe nanowires grown using the alloy nanoparticles.

Transport properties of the narrow SnTe nanowires were measured by fabricating nanodevices with 4-point contacts using standard e-beam lithography. For ohmic contacts, the contact areas were etched with Ar gas for two minutes and Cr/Au contacts were thermally deposited immediately after the etching. Figure 3b shows the resistance measurements of four SnTe nanowires as a function of temperature down to 1.7 K. The resistances were normalized to the room temperature resistance for comparison. The resistance values at room temperature were 55 Ω, 350 Ω, 154 Ω, 38 Ω for Device #1 (80 nm diameter), Device #2 (86 nm diameter), Device #3 (123 nm diameter), and Device #4 (300 nm), respectively. The gradual decrease in resistance at lower temperature indicates that the nanowires are heavily doped, which is common for SnTe due to Sn vacancies.[55] Interestingly, the residual resistance ratio (RRR) decreases systematically with decreasing diameter of the nanowires. As the RRR is a measure of electron scattering, thus a proxy for crystalline quality, the decrease in RRR indicates increased electron scattering. The increased scattering may be due to structural defects present in the nanowires, such as vacancies and surface stacking, and surface oxidation, which is more pronounced in transport measurements of nanoscale systems.[56]

The increased scattering is also reflected in the magnetoresistance curves measured at 1.7 K (Figure 3d). The magnetoresistance has the largest increase for the 300 nm-wide nanowire and gradually gets suppressed with decreasing diameter of the nanowire. This behavior agrees with the reduced RRR for narrower nanowires. Structural defects, such as Sn vacancies, may be affected by the dimensions of the SnTe nanowires or by the composition of the metal nanoparticles used to catalyze growth. In such case, the carrier density and mobility may change as a function of the nanowire diameter or the metal alloy composition. Due to the small diameter of the nanowires, Hall bar devices could not be fabricated to measure the carrier densities and Hall mobilities. As a rough estimate of the carrier density, we assume the measured magnetoresistance originates from a single carrier channel with one mobility value, and fit the magnetoresistance using the following equation: $\rho_{xx} = \frac{1}{2enu}(1 + \mu^2 B^2)$, where *e* is the electric charge, $n$ is the carrier density, $\mu$ is mobility, and $B$ is the magnetic field. From the fits to the magnetoresistance curves of Device #3 and #4, we estimate the carrier densities of $1 \times 10^{20}$ cm$^{-3}$ and $0.9 \times 10^{20}$ cm$^{-3}$ and mobility values of 4000 cm$^2$V$^{-1}$s$^{-1}$ and 2100 cm$^2$V$^{-1}$s$^{-1}$ for the two devices, respectively. These are on the lower end of the carrier densities we have previously measured in Hall bar devices of SnTe nanoplates and nanoribbons.[26] However we note that the mobility values seem very high, and the obtained values of the carrier densities and mobilities should be considered only as a rough estimate. We also note that the magnetoresistance of the 300 nm-wide nanowire does not follow a quadratic relation in which the resistance scales with B$^2$. Rather, the magnetoresistance appears to be more linear in high magnetic fields.

The transport measurements also show the ferroelectric transition of SnTe, which is accompanied by a structural transformation from its cubic structure at room temperature to the distorted rhombohedral structure.[57-58] The ferroelectric transition occurs at a temperature at which the resistance curve changes its slope with temperature. Figure 3c shows the derivatives of the resistance curves with respect to temperature, $d\rho/dT$, of Devices #1-4. The kinks observed in the resistance curves (Figure 3b, marked with arrows) are more clearly shown as sudden changes in the $d\rho/dT$ curves, which are marked by arrows in black and red (Figure 3c). We observe broad temperature windows over which $d\rho/dT$ curves change their slopes, which are denoted by shaded rectangular boxes in Figure 3c. Based on transport studies of bulk SnTe crystals,[57-58] we attribute the resistance change over the broad temperature window to the phase transition of the nanowires. The transition temperature is correlated with the carrier density; a higher carrier density results in a lower transition temperature.[58] In the case of SnTe nanowires, the broad temperature window for the phase transition as well as their dependence on the diameter of the nanowire may suggest additional effects, besides the carrier density, which determine the phase transition temperature, such as presence of surface defects and one-dimensional confinement effects.

To directly observe the structural phase transition and investigate the ferroelectric nature of the SnTe nanowires, we carried out *in situ* cryo-TEM experiments in which we cooled the narrow SnTe nanowires from room temperature to 12 K. At low temperature, dark bands appeared perpendicular to the nanowire [100] axis along the length of the nanowire, which were absent at room temperature (Figure 4a, 4b). These bands are not bend contours that would arise from local bending of the nanowire as swinging the electron beam did not change the locations of the dark bands. Additionally, the electron diffraction pattern taken at 25 K showed splitting of the (001) diffraction spots along the [100] axis (Figure 4d), which were single spots at room temperature (Figure 4c). Upon warming to room temperature, the dark bands as well as the splitting of the diffraction spots disappeared, indicating a reversible change. Thus, the *in situ* cryo-TEM data is indicative of a symmetry lowering structural phase transition at low temperature.

If the observed nanowire does, in fact, show similar ferroelectric order to that of bulk SnTe, the cubic rock-salt structure at room temperature (with α = 60° when described in the primitive rhombohedral cell) would undergo a rhombohedral distortion characterized by α ~59° and a polar displacement of Te along the [111] axis.[59-60] Within the pseudo-cubic unit cell, this results in a [111] oriented polarization, as illustrated in Figure 5b. While the ferroelectric domain structure of bulk SnTe is not known, rhombohedral $BiFeO_3$ and $BaTiO_3$ are well-studied systems which, similar to SnTe, have [111] polarization with a pseudo-cubic unit cell . In both $BiFeO_3$ and $BaTiO_3$, mechanical compatibility and charge neutrality across domain walls (DWs) favor the formation of 109° DWs on {100} or 71° DWs on {110} planes.[61-62] Figure 5b shows an atomic schematic of a 109° DW for SnTe, and Figure 5c shows a nanowire schematic with both a 109° and 71° DW. The dark bands observed with cryo-TEM imaging (Figure 4b) are consistent with either a 109° DW on (100) viewed edge on, or a 71° DW on (110) viewed in projection. To relate the proposed domain structures to the diffraction data (Figure 4d), we simulated the diffraction signals from both 109° and 71° DWs (see Experimental details). Both simulations qualitatively match two key aspects of the experimental data: first, splitting of (100) spots is absent owing to the invariant (100) or (110) DW plane, and secondly, (001) spots split along the [100] direction. However, both the 71° and 109° DW simulations show splitting of (101) type spots, which is absent in the experimental data. We found that the (101) type spot splitting can be removed if one of the domains is tilted ~1° about the [001] axis in simulations (Supplementary Figure S4). Thus, the discrepancy between the experimental and simulated data can be explained if there is a slight misorientation across the DW, possibly due to structural flexibility of unbounded nanowires. To summarize, the cryo-TEM data is fully consistent with ferroelectric SnTe with [111] polarization, and either 109° or 71° DWs.

**CONCLUSIONS**

We developed a reliable route to synthesize narrow SnTe nanowires with diameters under 100 nm using alloy nanoparticles. Experimentally, this was achieved by performing CVD growths twice. In the first growth, gold nanoparticles react with SnTe vapor to form the alloy nanoparticles. In the second growth, the alloy nanoparticles catalyze the growth of narrow SnTe nanowires. Transport measurements showed a size-dependent RRR and magnetoresistance, suggesting increased electron scattering due to surface oxidation and other factors. The diameter-dependent ferroelectric transition temperature suggests that the carrier density decreases for narrower nanowires, likely due to the removal of bulk electron carriers. *In situ* cryo-TEM experiments of SnTe nanowires showed the low temperature rhombohedral structure with the emergence of dark bands in the TEM image, suggesting domain boundaries between two rhombohedral domains. Simulation results suggest that both 109° and 71° domain walls are possible. The synthesized nanowires are important for the investigation of the nature of the superconductivity induced in these nanowires, which was recently suggested to have a pairing symmetry of $s \pm is$ via Josephson junction measurements. They are also important for investigations of Majorana bound states expected in this material.

**EXPERIMENTAL DETAILS**

*Structural Characterizations*: Scanning electron microscope (SEM) images and SEM-energy-dispersive X-ray spectroscopy (EDX) of SnTe nanowires were recorded using a Hitachi SU8230 cold field emission operated at 10 kV. Transmission electron microscopy (TEM) and TEM-EDX analysis were carried out using a 200 kV FEI Tecnai Osiris TEM. The *in situ* cryo-TEM experiments were performed using Gatan liquid-helium cooling holder (HCTDT 3010) and JEOL ARM200CF operated at 200 kV at Brookhaven National Laboratory.

*Device Fabrication*: For transport measurements, the synthesized nanowires were transferred to 300 nm $SiO_2$/Si substrates and spin-coated with copolymer layers of ~400 nm methyl methacrylate (MMA EL 8.5) and ~200 nm poly-methyl methacrylate (PMMA A3/A4) e-beam resist. For nanoscale patterning, a Raith EBPG 5000+ electron-beam lithography system was used. For ohmic contacts, the Oxford 100 Reactive Ion Etcher was used to remove surface oxides and resist residues from the exposed areas after the four-terminal device pattern was developed.

The samples were etched for 2 minutes at 40 W power with the Ar gas flow of 40 sccm. The samples were then loaded in a thermal evaporator for contacts formed by a wetting layer of 10 nm Cr followed by 100 – 300 nm Au. All of the electrical and magnetic field measurements were performed using the Quantum Design (QD) Dynacool system at a base temperature of 1.7 K. Four-probe resistance was measured with an AC lock-in technique at 17.78 Hz using source currents of 50 nA to 1μA. Magnetoresistance was measured in perpendicular magnetic fields up to 14 T at 1.7 K.

*Diffraction Simulations for In situ Cryo-TEM Data*: Electron diffraction simulations were performed with the multislice method using the Computem software package (Kirkland, Advanced Computing in Electron Microscopy, Springer US, 2010). The polar SnTe structure was taken from Ref.[60]. Initially, a polar SnTe crystal was aligned with the [001] crystal axis parallel to the simulation *z*-axis (the TEM optic axis). To form the 109° DW, the crystal was duplicated and rotated 90° about the simulation *z*-axis and then 90° about the simulation *x*-axis. The second crystal was then adjusted so that the (100) planes of the two crystals were coincident. Separate diffraction simulations were performed for the initial and rotated crystals, and then the simulation patterns were overlaid. A similar procedure was performed for the 71° DW, except the crystal was only rotated 90° about the *z*-axis, and then the (110) planes were made coincident. The crystals were disk-shaped, 16 nm thick and 40 nm in diameter, and were embedded in a 80 x 80 $nm^2$ simulation cell. The step size was 0.1 nm.

**FIGURES**

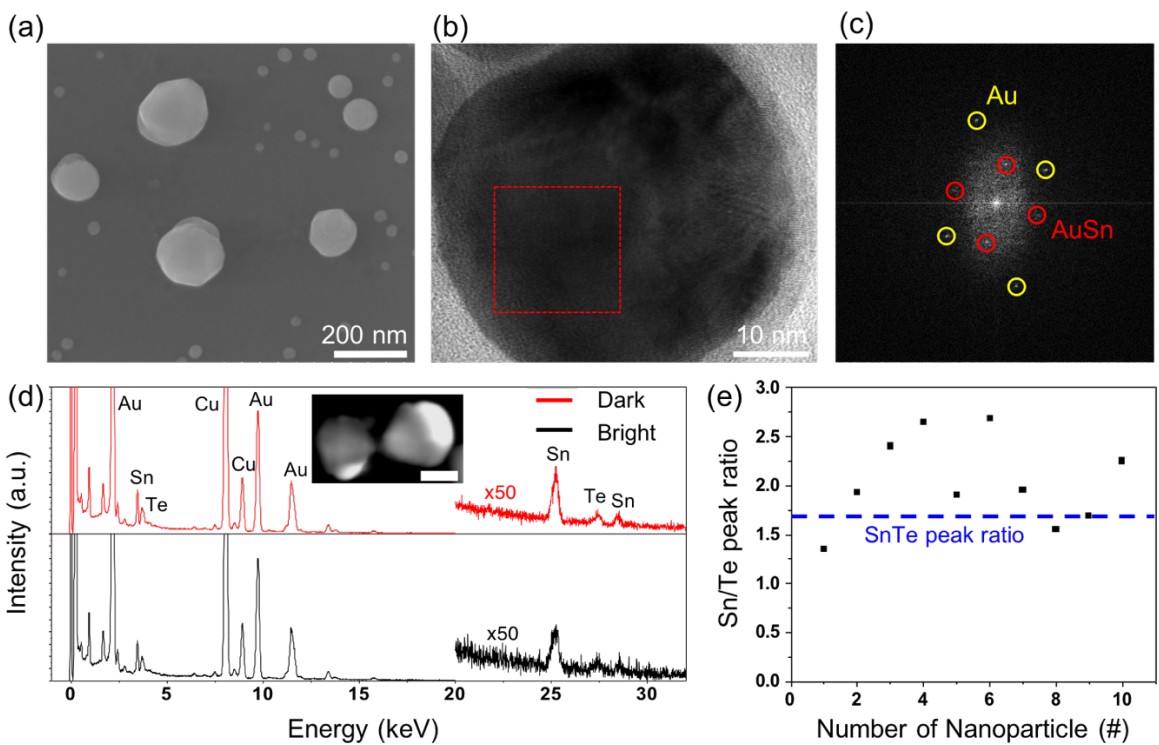

**Figure 1**. **Formation of Au-Sn-Te alloy nanoparticles.** (a) SEM image of a typical substrate after reacting Au nanoparticles with SnTe vapor. (b) TEM image of an alloy nanoparticle, showing lattice fringes. (c) A fast Fourier transform taken from the region of the TEM image (red box in (b)) shows diffraction spots that agree with diffraction spots of Au and AuSn. (d) EDX spectra of the nanoparticle show presence of Au, Sn, and Te. Cu is from the TEM grid. Inset shows the annular dark field STEM image of the nanoparticles; the EDX spectra were acquired from the particle on the right. (e) Composition analysis of the nanoparticles shows excess Sn in the alloy nanoparticles in comparison to SnTe, suggesting presence of AuSn and SnTe in the alloy nanoparticles.

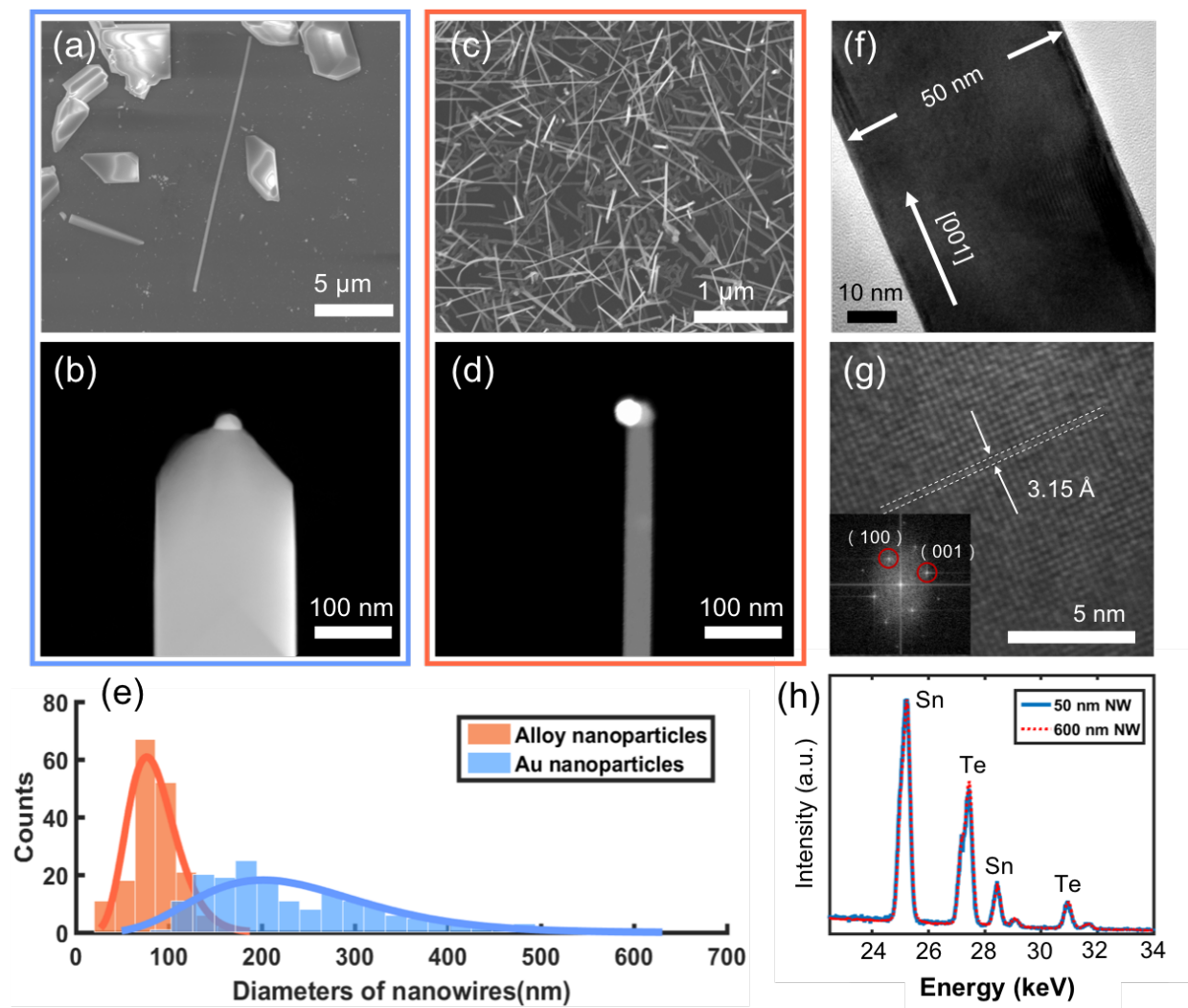

**Figure 2. Comparison of SnTe nanowires synthesized using Au versus alloy nanoparticles.** (a, b) SnTe nanowires grown using Au nanoparticles. (c,d) SnTe nanowires grown using the alloy nanoparticles. The width of the SnTe wire is much larger than the size of the metal particle in the case of using Au particles (b), while it is the same in the case of using alloy nanoparticles (d). (e) Histogram of diameters of SnTe nanowires synthesized using the alloy nanoparticles (orange) and the Au nanoparticles (blue). The average diameters are 85 nm and 240 nm, respectively. (f) TEM image of a 50 nm-wide SnTe nanowire. (g) High-resolution TEM image of the SnTe nanowire showing the expected cubic lattice at room temperature with the lattice spacing of 3.15 Å. Inset: Fast Fourier transform of the image, which shows the expected diffraction pattern of SnTe. (h) EDX spectra of a SnTe nanowire grown using a Au nanoparticle (600 nm NW) and the SnTe nanowire grown using the alloyed nanoparticle (50 nm NW). The spectra are identical, confirming the composition of the narrow nanowire is SnTe.

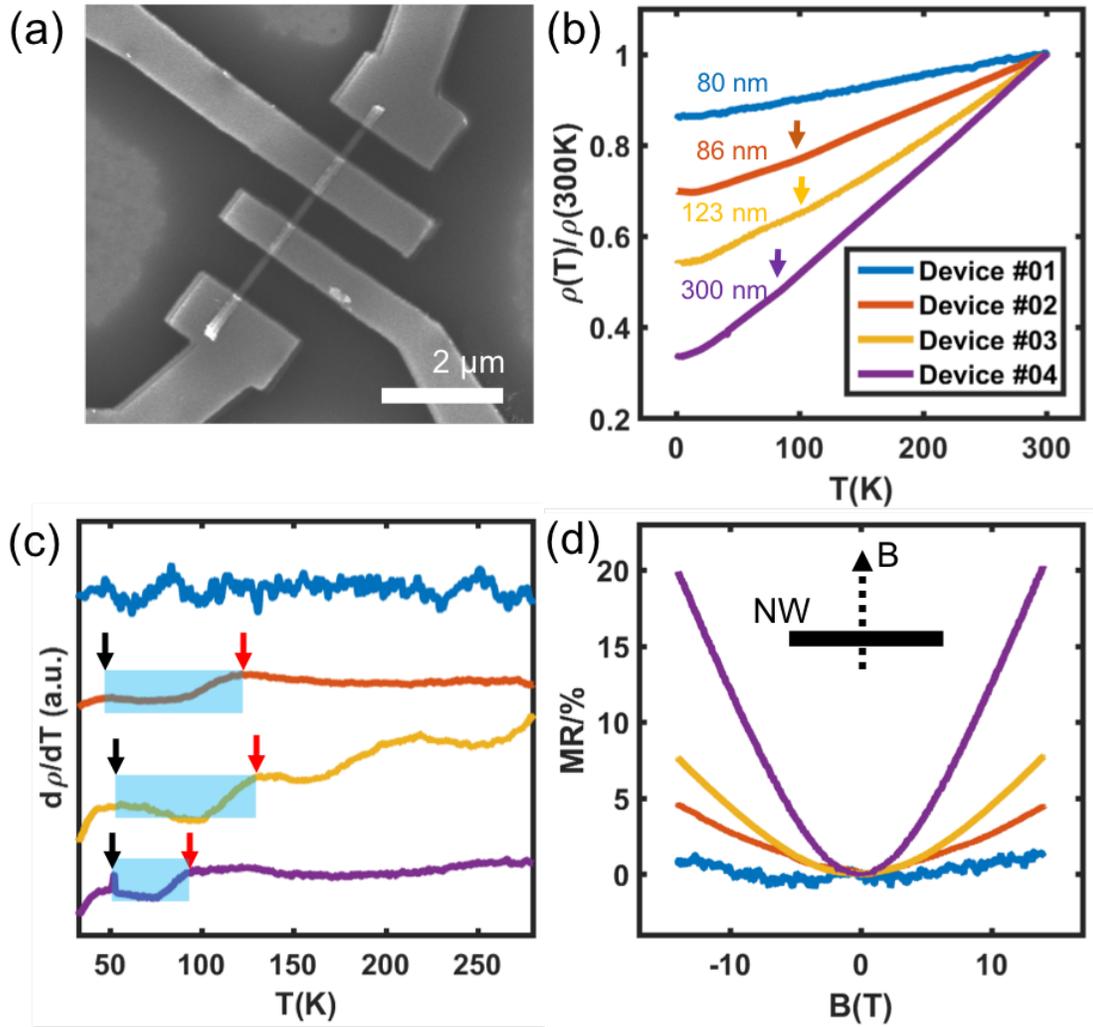

**Figure 3**. **Transport properties of narrow SnTe nanowires.** (a) SEM image of Device #1. (b) Temperature ($T$)-resistance ($\rho$) curves of SnTe nanowires of varying diameters: 80 nm (Device #1), 86 nm (Device #2), 123 nm (Device #3), and 300 nm (Device #4). The normalized resistance curves show slope changes at ~ 100 K, which are marked by arrows. (c) $d\rho/dT$ curves of the SnTe nanowires of varying diameters. Sudden changes in the $d\rho/dT$ reflect the slope changes in the $T$-$\rho$ curves shown in (a) better. The slope changes in the resistance indicate the structural phase transition. (d) Magnetic field ($B$)-resistance curves of the SnTe nanowires of varying diameters. Magnetoresistances (MR) are measured at T=1.7 K.

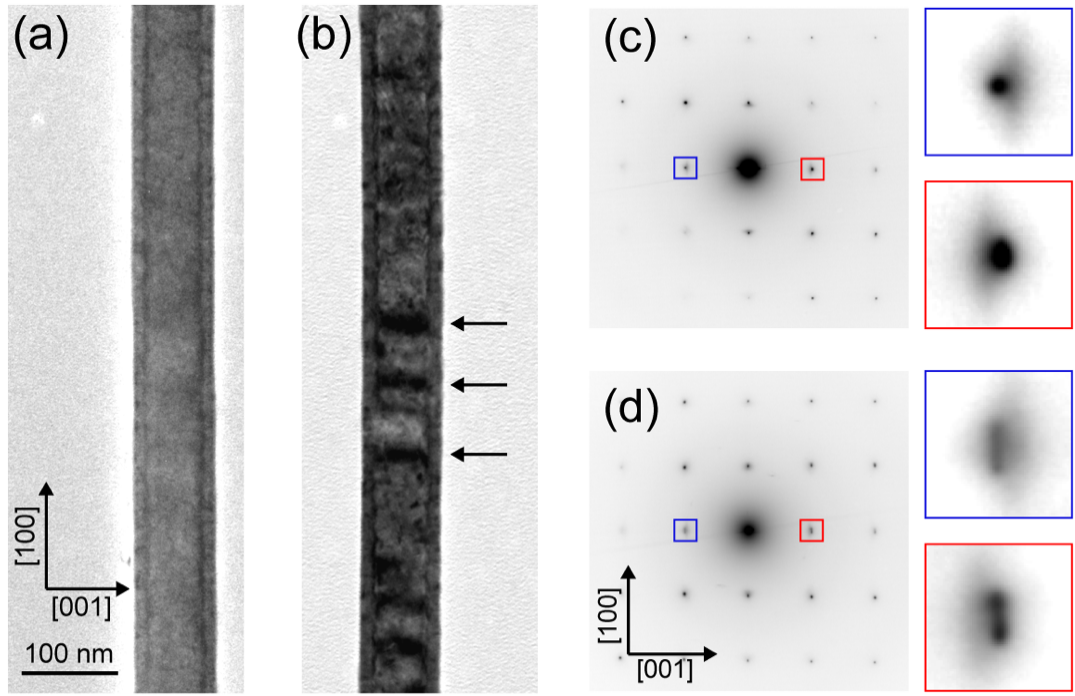

**Figure 4. Phase transition of SnTe nanowire by *in situ* cryo-TEM.** (a,b) TEM images of a SnTe nanowire at room temperature (a) and at 25 K (b). Dark bands appear along the nanowire at 25 K (marked by arrows in b), which are not present at room temperature. (c,d) Selected area electron diffraction patterns of the SnTe nanowire at room temperature (c) and at 25 K (d). In both c and d, the insets show magnified (001) type spots. Spot splitting is observed at 25 K but not at room temperature.

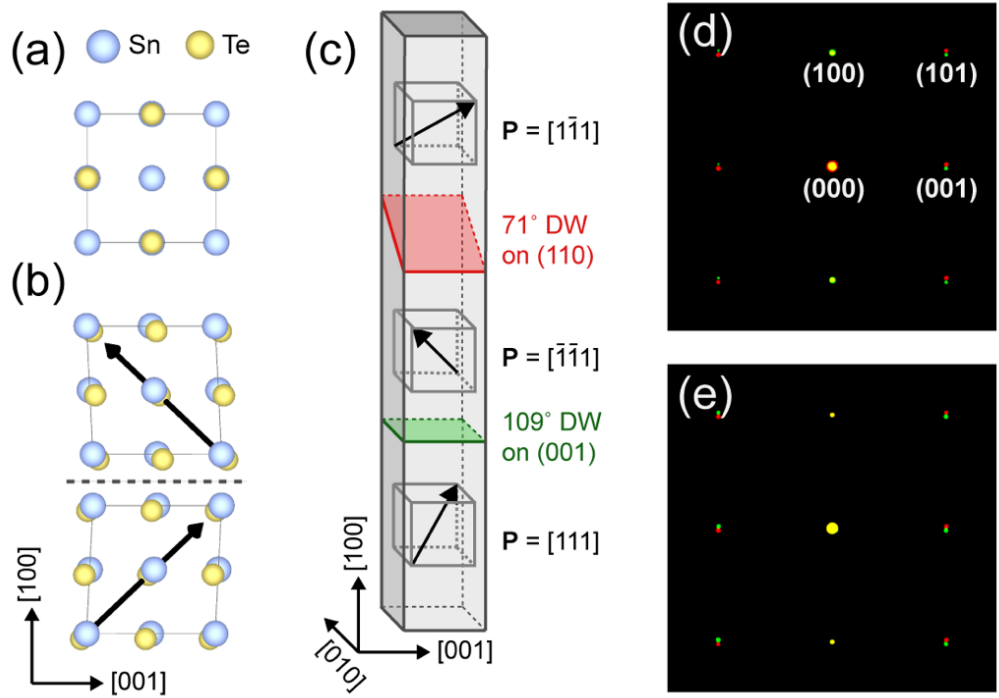

**Figure 5. Ferroelectric domain structures in SnTe nanowire at low temperature.** (a) Schematic showing the cubic SnTe unit cell at room temperature. (b) Schematic showing the polar SnTe pseudo-cubic unit cell at low temperature. The polar distortions were magnified for clarity. Two unit cells are shown to demonstrate a 109° DW on the (100) plane, indicated with the dotted line. The polarization changes from [111] to [1-1-1]. (c) Schematic nanowire with three domains, showing a 71° DW on (110) in red and a 109° DW on (100) in green. The rhombohedral unit cell distortion and its effect on the nanowire geometry is suppressed in this schematic for clarity. (d) Simulated electron diffraction of a 71° DW. **(e)** Simulated electron diffraction of a 109° DW. In both d and e, the different color spots represent diffraction spots from different domains.

## ASSOCIATED CONTENT

**Supporting Information**
The supporting information is available free of charge on the ACS Publication website http://pubs.acs.org.

## AUTHOR INFORMATION

**Corresponding Author**
*E-mail: judy.cha@yale.edu

**Notes**
The authors declare no competing financial interest.

**Author Contributions**
The manuscript was written through contributions of all authors. All authors have given approval to the final version of the manuscript.

## ACKNOWLEDGMENTS

Synthesis of the narrow SnTe nanowires was supported by NSF DMR-1743896. Transport measurements of the synthesized nanowires were supported by the Department of Energy Basic Energy Sciences DE-SC0014476. C.J.T. and J.R.W. acknowledge support from NSF DMR-1743913. The work at Brookhaven National Laboratory was supported by the Materials Science and Engineering Divisions, Office of Basic Energy Sciences of the U.S. Department of Energy under Contract No. DESC0012704.

## REFERENCES

1.     Hamdou, B.; Gooth, J.; Dorn, A.; Pippel, E.; Nielsch, K., Surface state dominated transport in topological insulator $Bi_2Te_3$ nanowires. *Applied Physics Letters* **2013**, *103* (19), 193107.
2.     Hamdou, B.; Gooth, J.; Dorn, A.; Pippel, E.; Nielsch, K., Aharonov-Bohm oscillations and weak antilocalization in topological insulator $Sb_2Te_3$ nanowires. *Applied Physics Letters* **2013**, *102* (22), 223110.
3.     Hong, S. S.; Zhang, Y.; Cha, J. J.; Qi, X.-L.; Cui, Y., One-Dimensional Helical Transport in Topological Insulator Nanowire Interferometers. *Nano Letters* **2014**, *14* (5), 2815-2821.
4.     Cho, S.; Dellabetta, B.; Zhong, R.; Schneeloch, J.; Liu, T.; Gu, G.; Gilbert, M. J.; Mason, N., Aharonov–Bohm oscillations in a quasi-ballistic three-dimensional topological insulator nanowire. *Nature Communications* **2015**, *6* (1), 7634.
5.     Peng, H.; Lai, K.; Kong, D.; Meister, S.; Chen, Y.; Qi, X.-L.; Zhang, S.-C.; Shen, Z.-X.; Cui, Y., Aharonov–Bohm interference in topological insulator nanoribbons. *Nature Materials* **2010**, *9* (3), 225.

6. Jauregui, L. A.; Pettes, M. T.; Rokhinson, L. P.; Shi, L.; Chen, Y. P., Magnetic field-induced helical mode and topological transitions in a topological insulator nanoribbon. *Nature Nanotechnology* **2016,** *11*, 345.
7. Zhang, C.; Zhang, Y.; Yuan, X.; Lu, S.; Zhang, J.; Narayan, A.; Liu, Y.; Zhang, H.; Ni, Z.; Liu, R., Quantum Hall effect based on Weyl orbits in $Cd_3As_2$. *Nature* **2019,** *565* (7739), 331.
8. Kitaev, A. Y., Unpaired Majorana fermions in quantum wires. *Physics-Uspekhi* **2001,** *44* (10S), 131.
9. Ivanov, D. A., Non-Abelian statistics of half-quantum vortices in p-wave superconductors. *Physical Review letters* **2001,** *86* (2), 268.
10. Alicea, J., New directions in the pursuit of Majorana fermions in solid state systems. *Reports on progress in physics* **2012,** *75* (7), 076501.
11. Liu, P.; Williams, J. R.; Cha, J. J., Topological nanomaterials. *Nature Reviews Materials* **2019,** *4* (7), 479-496.
12. Lutchyn, R. t.; Bakkers, E.; Kouwenhoven, L. P.; Krogstrup, P.; Marcus, C.; Oreg, Y., Majorana zero modes in superconductor–semiconductor heterostructures. *Nature Reviews Materials* **2018,** *3* (5), 52-68.
13. Tanaka, Y.; Ren, Z.; Sato, T.; Nakayama, K.; Souma, S.; Takahashi, T.; Segawa, K.; Ando, Y., Experimental realization of a topological crystalline insulator in SnTe. *Nature Physics* **2012,** *8* (11), 800-803.
14. Hsieh, T. H.; Lin, H.; Liu, J.; Duan, W.; Bansil, A.; Fu, L., Topological crystalline insulators in the SnTe material class. *Nature Communications* **2012,** *3* (1), 982.
15. Chang, K.; Liu, J.; Lin, H.; Wang, N.; Zhao, K.; Zhang, A.; Jin, F.; Zhong, Y.; Hu, X.; Duan, W.; Zhang, Q.; Fu, L.; Xue, Q.-K.; Chen, X.; Ji, S.-H., Discovery of robust in-plane ferroelectricity in atomic-thick SnTe. *Science* **2016,** *353* (6296), 274-278.
16. Yan, C.; Liu, J.; Zang, Y.; Wang, J.; Wang, Z.; Wang, P.; Zhang, Z.-D.; Wang, L.; Ma, X.; Ji, S.; He, K.; Fu, L.; Duan, W.; Xue, Q.-K.; Chen, X., Experimental Observation of Dirac-like Surface States and Topological Phase Transition in $Pb_{1-x}Sn_xTe$ (111) Films. *Physical Review Letters* **2014,** *112* (18), 186801.
17. Kumaravadivel, P.; Pan, G. A.; Zhou, Y.; Xie, Y.; Liu, P.; Cha, J. J., Synthesis and superconductivity of In-doped SnTe nanostructures. *APL Materials* **2017,** *5* (7), 076110.
18. Trimble, C.; Wei, M.; Yuan, N.; Kalantre, S.; Liu, P.; Cha, J.; Fu, L.; Williams, J., Josephson Detection of Time Reversal Symmetry Breaking Superconductivity in SnTe Nanowires. *arXiv preprint arXiv:1907.04199* **2019**.
19. Sasaki, S.; Ando, Y., Superconducting $Sn_{1-x}In_xTe$ Nanoplates. *Crystal Growth & Design* **2015,** *15* (6), 2748-2752.
20. Novak, M.; Sasaki, S.; Kriener, M.; Segawa, K.; Ando, Y., Unusual nature of fully gapped superconductivity in In-doped SnTe. *Physical Review B* **2013,** *88* (14), 140502.
21. Xu, E. Z.; Li, Z.; Martinez, J. A.; Sinitsyn, N.; Htoon, H.; Li, N.; Swartzentruber, B.; Hollingsworth, J. A.; Wang, J.; Zhang, S. X., Diameter dependent thermoelectric properties of individual SnTe nanowires. *Nanoscale* **2015,** *7* (7), 2869-2876.
22. Safdar, M.; Wang, Q.; Mirza, M.; Wang, Z.; Xu, K.; He, J., Topological Surface Transport Properties of Single-Crystalline SnTe Nanowire. *Nano Letters* **2013,** *13* (11), 5344-5349.
23. Kong, D.; Randel, J. C.; Peng, H.; Cha, J. J.; Meister, S.; Lai, K.; Chen, Y.; Shen, Z.-X.; Manoharan, H. C.; Cui, Y., Topological insulator nanowires and nanoribbons. *Nano Letters* **2009,** *10* (1), 329-333.
24. Naylor, C. H.; Parkin, W. M.; Ping, J.; Gao, Z.; Zhou, Y. R.; Kim, Y.; Streller, F.; Carpick, R. W.; Rappe, A. M.; Drndic, M., Monolayer single-crystal 1T′-$MoTe_2$ grown by chemical vapor deposition exhibits weak antilocalization effect. *Nano Letters* **2016,** *16* (7), 4297-4304.

25. Li, Z.; Shao, S.; Li, N.; McCall, K.; Wang, J.; Zhang, S. X., Single Crystalline Nanostructures of Topological Crystalline Insulator SnTe with Distinct Facets and Morphologies. *Nano Letters* **2013,** *13* (11), 5443-5448.
26. Shen, J.; Jung, Y.; Disa, A. S.; Walker, F. J.; Ahn, C. H.; Cha, J. J., Synthesis of SnTe nanoplates with {100} and {111} surfaces. *Nano Letters* **2014,** *14* (7), 4183-4188.
27. Liu, P.; Xie, Y.; Miller, E.; Ebine, Y.; Kumaravadivel, P.; Sohn, S.; Cha, J. J., Dislocation-driven SnTe surface defects during chemical vapor deposition growth. *Journal of Physics and Chemistry of Solids* **2019,** *128*, 351-359.
28. Zhang, E.; Liu, Y.; Wang, W.; Zhang, C.; Zhou, P.; Chen, Z.-G.; Zou, J.; Xiu, F., Magnetotransport properties of $Cd_3As_2$ nanostructures. *ACS Nano* **2015,** *9* (9), 8843-8850.
29. Zhang, K.; Pan, H.; Zhang, M.; Wei, Z.; Gao, M.; Song, F.; Wang, X.; Zhang, R., Controllable synthesis and magnetotransport properties of $Cd_3As_2$ Dirac semimetal nanostructures. *RSC Advances* **2017,** *7* (29), 17689-17696.
30. Wagner, R.; Ellis, W., Vapor‐liquid‐solid mechanism of single crystal growth. *Applied Physics Letters* **1964,** *4* (5), 89-90.
31. Cui, Y.; Lauhon, L. J.; Gudiksen, M. S.; Wang, J.; Lieber, C. M., Diameter-controlled synthesis of single-crystal silicon nanowires. *Applied Physics Letters* **2001,** *78* (15), 2214-2216.
32. Dick, K. A., A review of nanowire growth promoted by alloys and non-alloying elements with emphasis on Au-assisted III–V nanowires. *progress in Crystal Growth and Characterization of Materials* **2008,** *54* (3-4), 138-173.
33. Güniat, L.; Caroff, P.; Fontcuberta i Morral, A., Vapor phase growth of semiconductor nanowires: Key developments and open questions. *Chemical Reviews* **2019,** *119* (15), 8958-8971.
34. Safdar, M.; Wang, Q.; Mirza, M.; Wang, Z.; He, J., Crystal Shape Engineering of Topological Crystalline Insulator SnTe Microcrystals and Nanowires with Huge Thermal Activation Energy Gap. *Crystal Growth & Design* **2014,** *14* (5), 2502-2509.
35. Atherton, S.; Steele, B.; Sasaki, S., Unexpected Au Alloying in Tailoring In-Doped SnTe Nanostructures with Gold Nanoparticles. *Crystals* **2017,** *7* (3), 78.
36. Sadowski, J.; Dziawa, P.; Kaleta, A.; Kurowska, B.; Reszka, A.; Story, T.; Kret, S., Defect-free SnTe topological crystalline insulator nanowires grown by molecular beam epitaxy on graphene. *Nanoscale* **2018,** *10* (44), 20772-20778.
37. Saghir, M.; Lees, M. R.; York, S. J.; Balakrishnan, G., Synthesis and Characterization of Nanomaterials of the Topological Crystalline Insulator SnTe. *Crystal Growth & Design* **2014,** *14* (4), 2009-2013.
38. Frank, F. C., Supercooling of liquids. *Proceedings of the Royal Society of London. Series A. Mathematical and Physical Sciences* **1952,** *215* (1120), 43-46.
39. Liu, X., Heterogeneous nucleation or homogeneous nucleation? *The Journal of Chemical Physics* **2000,** *112* (22), 9949-9955.
40. Kolasinski, K. W., Catalytic growth of nanowires: vapor–liquid–solid, vapor–solid–solid, solution–liquid–solid and solid–liquid–solid growth. *Current Opinion in Solid State and Materials Science* **2006,** *10* (3-4), 182-191.
41. Sun, Y.-L.; Matsumura, R.; Jevasuwan, W.; Fukata, N., Au–Sn Catalyzed Growth of $Ge_{1-x}Sn_x$ Nanowires: Growth Direction, Crystallinity, and Sn Incorporation. *Nano Letters* **2019,** *19* (9), 6270-6277.
42. Maliakkal, C. B.; Jacobsson, D.; Tornberg, M.; Persson, A. R.; Johansson, J.; Wallenberg, R.; Dick, K. A., In situ analysis of catalyst composition during gold catalyzed GaAs nanowire growth. *Nature Communications* **2019,** *10* (1), 4577.
43. Dubrovskii, V.; Sibirev, N., General form of the dependences of nanowire growth rate on the nanowire radius. *Journal of Crystal Growth* **2007,** *304* (2), 504-513.

44. Johansson, J.; Svensson, C. P. T.; Mårtensson, T.; Samuelson, L.; Seifert, W., Mass transport model for semiconductor nanowire growth. *The Journal of Physical Chemistry B* **2005,** *109* (28), 13567-13571.
45. Dayeh, S. A.; Picraux, S. T., Direct Observation of Nanoscale Size Effects in Ge Semiconductor Nanowire Growth. *Nano Letters* **2010,** *10* (10), 4032-4039.
46. Givargizov, E. I., Fundamental aspects of VLS growth. *Journal of Crystal Growth* **1975,** *31*, 20-30.
47. Shen, Y.; Chen, R.; Yu, X.; Wang, Q.; Jungjohann, K. L.; Dayeh, S. A.; Wu, T., Gibbs–Thomson Effect in Planar Nanowires: Orientation and Doping Modulated Growth. *Nano Letters* **2016,** *16* (7), 4158-4165.
48. Tornberg, M.; Jacobsson, D.; Persson, A. R.; Wallenberg, R.; Dick, K. A.; Kodambaka, S., Kinetics of Au–Ga Droplet Mediated Decomposition of GaAs Nanowires. *Nano Letters* **2019,** *19* (6), 3498-3504.
49. Gao, H.; Sun, Q.; Lysevych, M.; Tan, H. H.; Jagadish, C.; Zou, J., Effect of Sn Addition on Epitaxial GaAs Nanowire Grown at Different Temperatures in Metal–Organic Chemical Vapor Deposition. *Crystal Growth & Design* **2019,** *19* (9), 5314-5319.
50. Mårtensson, E. K.; Whiticar, A. M.; de la Mata, M.; Zamani, R. R.; Johansson, J.; Nygård, J.; Dick, K. A.; Bolinsson, J., Understanding GaAs Nanowire Growth in the Ag–Au Seed Materials System. *Crystal Growth & Design* **2018,** *18* (11), 6702-6712.
51. Dayeh, S. A.; Yu, E. T.; Wang, D., Surface Diffusion and Substrate−Nanowire Adatom Exchange in InAs Nanowire Growth. *Nano Letters* **2009,** *9* (5), 1967-1972.
52. Yan, X.; Li, B.; Zhang, X.; Ren, X., Growth of pure wurtzite InAs nanowires over a wide diameter range. *Applied Surface Science* **2018,** *458*, 269-272.
53. Assali, S.; Gagliano, L.; Oliveira, D. S.; Verheijen, M. A.; Plissard, S. R.; Feiner, L. F.; Bakkers, E. P. A. M., Exploring Crystal Phase Switching in GaP Nanowires. *Nano Letters* **2015,** *15* (12), 8062-8069.
54. Halder, N. N.; Kelrich, A.; Cohen, S.; Ritter, D., Pure wurtzite GaP nanowires grown on zincblende GaP substrates by selective area vapor liquid solid epitaxy. *Nanotechnology* **2017,** *28* (46), 465603.
55. Wang, N.; West, D.; Liu, J.; Li, J.; Yan, Q.; Gu, B.-L.; Zhang, S.; Duan, W., Microscopic origin of the p-type conductivity of the topological crystalline insulator SnTe and the effect of Pb alloying. *Physical Review B* **2014,** *89* (4), 045142.
56. Woods, J. M.; Hynek, D.; Liu, P.; Li, M.; Cha, J. J., Synthesis of $WTe_2$ Nanowires with Increased Electron Scattering. *ACS Nano* **2019,** *13* (6), 6455-6460.
57. Katayama, S.; Mills, D., Theory of anomalous resistivity associated with structural phase transitions in IV-VI compounds. *Physical Review B* **1980,** *22* (1), 336.
58. Kobayashi, K. L. I.; Kato, Y.; Katayama, Y.; Komatsubara, K. F., Carrier-Concentration-Dependent Phase Transition in SnTe. *Physical Review Letters* **1976,** *37* (12), 772-774.
59. Rabe, K. M.; Joannopoulos, J. D., Ab initio relativistic pseudopotential study of the zero-temperature structural properties of SnTe and PbTe. *Physical Review B* **1985,** *32* (4), 2302-2314.
60. Plekhanov, E.; Barone, P.; Di Sante, D.; Picozzi, S., Engineering relativistic effects in ferroelectric SnTe. *Physical Review B* **2014,** *90* (16), 161108.
61. Marton, P.; Rychetsky, I.; Hlinka, J., Domain walls of ferroelectric $BaTiO_3$ within the Ginzburg-Landau-Devonshire phenomenological model. *Physical Review B* **2010,** *81* (14), 144125.
62. Silva, J.; Reyes, A.; Esparza, H.; Camacho, H.; Fuentes, L., $BiFeO_3$: A Review on Synthesis, Doping and Crystal Structure. *Integrated Ferroelectrics* **2011,** *126* (1), 47-59.